\documentclass{aa} 
\usepackage[varg]{txfonts}    
\begin{document}


\title{Principal component analysis of sunspot cycle shape }

\author{Jouni Takalo
\and
Kalevi Mursula
}

\institute{ReSoLVE Centre of Excellence, Space Climate research unit, University of Oulu,
POB 3000, FIN-90014, Oulu, Finland\\
\email{jouni.j.takalo@oulu.fi}; \email{kalevi.mursula@oulu.fi}
}

\date{Received: }

\abstract {}
{We study the shape of sunspot cycles using the Wolf sunspot numbers and group sunspot numbers of solar cycles 1-23. We determine the most typical "model" cycles and the most asymmetric cycles, and test the validity of the two Waldmeier rules: the anti-correlation between cycle height and the length of its ascending phase (rule 1), and between cycle height and the length of the preceding cycle (rule 2).} 
{We applied the principal component analysis to sunspot cycles and studied the first two components, which describe the average cycle shape and cycle asymmetry, respectively. We also calculated their autocorrelation in order to study their recurrence properties.} 
{The best model cycles for Wolf numbers are SC12, SC14, and SC16, the successive even cycles from a long period of rather low overall solar activity. We find that the model cycles in eight different analyses using both sunspot series are almost exclusively even cycles. Correspondingly, the most asymmetric cycles are odd cycles.
We find that both Waldmeier rules are valid for the whole Wolf number series of 23 cycles. Waldmeier rule 2 is also valid for group number series although its significance is weaker. Waldmeier rule 1 is not significant for the original group number series, but becomes significant for the proxy series. 
For separate centuries, Waldmeier rules are not always valid for Wolf numbers and very rarely for group numbers. }
{The preference of even cycles as model cycles supports the Gnevyshev-Ohl rule and the related 22-year alternation of cycle amplitudes and intensities, with even cycles on average being 10-15\% lower than odd cycles. Our results also offer a new interpretation for the Gnevyshev gap. In addition to being a local depression of solar activity, the Gnevyshev gap is a separatrix that divides cycles into two parts whose relative intensities determine the cycle asymmetry. The Gnevyshev gap is the zero value time of PC2, located approximately 33-42\% into the cycle after its start.}

\keywords{Sun: sunspots, Sun: activity, Sun: general, methods: data analysis}

\titlerunning{Principal component analysis of sunspot cycle shape}
\authorrunning{Takalo and Mursula}

\maketitle

\section{Introduction}

It has been known for almost two hundred years that the occurrence of sunspots is, at least most of the time, cyclic, although not strictly periodic. The length of the sunspot cycle (SC) has varied between 9.0 and 13.7 years. The shape of the sunspot cycle has also changed. Waldmeier noticed the asymmetry of sunspot cycles, with the ascending phase being typically shorter than the declining phase, and that there is an anti-correlation between cycle amplitude and the length of the ascending phase of the cycle \citep{Waldmeier_1935, Waldmeier_1939} (original Waldmeier effect; here Waldmeier rule 1). The shape of the sunspot cycle has been fitted by several different functions and methods. \cite {Nordemann_1992_1} and \cite {Nordemann_1992_2} fitted  separate exponential functions to the ascending and declining phases of the sunspot cycle using three parameters for both phases. They concluded that there are periodicities along which the cycle shape repeats the same pattern, the dominating periods being 59 and 86 years. \cite{Elling_1992} found that the shape of the full sunspot cycle could be presented by a modulated F-distribution density function with five parameters. \cite{Hathaway_1994} suggested a function similar to that of \cite{Stewart_1938}, but with a cubic power law for the ascending phase and a gaussian function for the declining phase, which were made to smoothly join together at sunspot minima and maxima. More recently, \cite{Volobuev_2009} optimized a similar function with two parameters, which reduced to a one-parameter fit if the times of sunspot minima were a priori known. \cite{Du_2011} derived the shape of the sunspot cycle from a modified gaussian function with four parameters.

In the aforementioned paper, \cite{Hathaway_1994} presented a modified Waldmeier effect (here Waldmeier rule 2), that is, the anticorrelation between cycle amplitude and the length of the preceding cycle, and found a weak anticorrelation for sunspot cycles 2-18, supporting rule 2. Later \cite{Hathaway_2002} studied group sunspot numbers (GSN) and compared them to Z\"urich sunspot numbers (SSN). They concluded that Waldmeier rules are stronger for SSN than for GSN. Recently, \cite{Aparicio_2012} also verified that Waldmeier rules are more apparent in SSN than in GSN.

Some papers have studied the Waldmeier rules using sunspot area \citep{Dikpati_2008, Karak_2011} separately for odd and even numbered cycles \citep{Dikpati_2008} or separately for small and large sunspot groups \citep{Javaraiah_2012}. Dikpati et al. (2008) concluded that Waldmeier rule 1 seems to be valid only for even numbered sunspot cycles. For all odd cycles since cycle 13, the rising time was found to be independent of cycle height. They also reported that Waldmeier rule 1 is not valid for sunspot areas, since the epochs of the maxima of sunspot numbers and sunspot areas are different for some cycles. However, Karak and Choudhuri (2011) defined the rising time as the time during which the activity level changes from 0.2P to 0.8P, where P is the cycle maximum, and found that Waldmeier rule 1 applies also to sunspot areas using this definition. Javaraiah (2012) studied Waldmeier rule 1 in the annual counts of small, medium-sized, and large sunspot groups. He found that for many cycles the positions of the maxima of the small, medium-sized, and large groups are different and also considerably differ from the corresponding SSN maxima. 

In this study we use the principal component analysis (PCA) method to calculate the average shape of the sunspot cycle for the last 23 solar cycles using SSN and GSN. We also calculate the mean shape of the sunspot cycle in the eighteenth, nineteenth, and twentieth centuries separately. We compare the time series constructed from the two main principal components (PC1 and PC2) with the original time series, and study the validity of the Waldmeier rules for them. This paper is organized as follows. Section 2 presents the data and methods used in this study. In Sect. 3 we present the results of PC analysis for the cycle shape using sunspot numbers and group sunspot numbers for cycles 1-23. In Sect. 4 we compare the shapes of eighteenth, nineteenth and twentieth century cycles. In Sect. 5 we present statistics of the original sunspot indices and their PC proxies and study the validity of the Waldmeier rules. We give our conclusions in Sect. 6.

\section{Data and methods}

\subsection{Sunspot indices}

Rudolf Wolf collected the Z\"urich series of sunspot numbers (SSN), which started in 1749. The first complete sunspot cycle included in SSN started in March 1755. Wolf started the numbering of sunspot cycles from this cycle and this numbering is still in use. The initial sunspot number series (here called SSN1) was reconstructed at the Z\"urich Observatory until 1980, and has been reconstructed at the Royal Observatory of Belgium since 1981. Following the change of the reconstruction method in 1981, the current version of the SSN series is called the international sunspot number (ISN). Recently the ISN series was modified to a version 2.0 that is supposed to present a preliminary correction of the known inhomogeneities in the SSN1 series \citep{Clette_2014}. We use the monthly indices of SSN1 in this paper, but we have verified that using SSN2 gives very similar results to SSN1. The dates of the sunspot minima and the cycle lengths for SSN1 are shown in Table 1.

\begin{table}
\begin{center}
\small
\caption{Sunspot cycle lengths (in years) and dates (fractional years, and year and month) of (starting) sunspot minima for SSN1 \citep{NGDC_2013}.}
\begin{tabular}{  c  c  c  c }
\hline
   Cycle & Frac. year    &Year\&month     &Cycle  \\
            number     &of minimum     & of minimum     &    length \\
        \hline 
1      &1755.2   &1775 March  & 11.3  \\
2      &1766.5   &1766 June  & 9.0  \\
3      &1775.5   &1775 June  & 9.2  \\
4      &1784.7   &1784 September  &	13.7  \\
5      &1798.4   &1798 May  & 12.2  \\
6      &1810.6   &1810 August  & 12.7  \\
7      &1823.3   &1823 April & 10.6  \\
8      &1833.9   &1833 November & 9.6  \\
9      & 1843.5  &1843 July  & 12.5  \\
10     & 1856.0  &1855 December  & 11.2  \\
11     & 1867.2  &1867 March  & 11.8  \\
12     & 1879.0  &1878 December  & 10.6  \\
13     & 1889.6  &1889 August  & 12.1  \\
14     & 1901.7  &1901 September  & 11.8  \\
15     & 1913.5  &1913 July  & 10.1  \\
16     & 1923.6  &1923 August & 10.1  \\
17     & 1933.7  &1933 September  & 10.4  \\
18     & 1944.1  &1944 February  & 10.2  \\
19     & 1954.3  &1954 April  & 10.5  \\
20     & 1964.8  &1964 October  & 11.7  \\
21     & 1976.5  &1976 June & 10.2  \\
22     & 1986.7  &1986 September  & 10.1  \\
23     & 1996.8  &1996 October  & 12.2  \\
24     & 2009.0  &2008 December  &       \\

\hline
\end{tabular}
\end{center}
\end{table}

The group sunspot number (GSN) starts as early as 1610 \citep{Hoyt_1998}, but has scarce coverage until solar cycle 1 (SC1), and some missing monthly values still up to SC5 (which were partly filled by geomagnetic proxy data in SSN). The original GSN series ends in 1995, thus missing SC23. We have filled in the gaps in monthly GSN data in SC1-SC4 by using linear interpolation. We have also continued the GSN series by using the recently published group sunspot number time series \citep{Chatzistergos_2017} and adjusted it to the level of average GSN time series in SC15-SC22. It seems that the minima of the GSN index are not always the same as in SSN1. Therefore we have defined the minima of the GSN data using Gleissberg-smoothed (smoothing over 13 months) GSN time series. The dates of GSN minima and their difference to SSN1 minima are shown in Table 2.

\begin{table}
\begin{center}
\small
\caption{Dates (fractional years, and year and month) of (starting) minima of GSN cycles, GSN cycle lengths, and their difference to SSN1 minima (in months).}
\begin{tabular}{  c  c  c  c  c }
\hline
    Cycle & Frac. year    &Year\&month     &Cycle  &Diff. to \\
    number     &of minimum     & of minimum     &length & SSN1 min\\
        \hline 
1      &1755.2   &1775 March & 11.2    & 0 \\
2      &1766.4   &1766 May & 9.1  & -1 \\
3      &1775.5   &1775 June & 9.0  & 0 \\ 
4      &1784.5   &1784 July & 14.2  & -2 \\
5      &1798.7   &1798 September & 11.8 & +4  \\
6      &1810.5   &1810 July & 12.8 & -1 \\
7      &1823.3   &1823 April & 10.6 &  0  \\
8      &1833.9   &1833 November & 9.7 &  0 \\
9      & 1843.6  &1843 August & 12.5 & +1  \\
10     & 1856.1  &1856 February  & 11.2 & +2  \\
11     & 1867.3  &1867 April  & 11.7 & +1  \\
12     & 1879.0  &1879 January & 10.9  &  +1 \\
13     & 1889.9  &1889 November & 12.1 & +3  \\
14     & 1902.0  &1901 December & 11.6  & +3 \\
15     & 1913.6  &1913 August & 10.2  & +1  \\
16     & 1923.8  &1923 October & 10.1 & +2 \\
17     & 1933.9  &1933 November & 10.4  & +2 \\
18     & 1944.3  &1944 April & 10.0 & +2 \\
19     & 1954.3  &1954 April & 10.4 &  0  \\
20     & 1964.7  &1964 September & 11.8  & -1  \\
21     & 1976.5  &1976 June & 10.0 &  0 \\
22     & 1986.5  &1986 June & 10.2 & -3  \\
23     & 1996.7  &1996 September & 12.4 & -1  \\
24     & 2009.1  &2009 February  &   &  +2   \\

\hline
\end{tabular}
\end{center}
\end{table}

\subsection{PCA method} 
Principal component analysis is a useful tool in many fields of science including chemometrics \citep{Bro_2014}, data compression \citep{Kumar_2008}, and information extraction \citep{Hannachi_2007}. For a large number of correlated variables, PCA finds combinations of a few uncorrelated variables that describe the majority of variability in the data. PCA has earlier been applied, for example, to studies of the geomagnetic field \citep{Bhattacharyya_2015}, geomagnetic activity \citep{Holappa_2014_2, Holappa_2014_1}, the ionosphere \citep{Lin_2012}, and the solar background magnetic field \citep{Zharkova_2012, Zharkova_2016}. As far as we know, PCA has not yet been applied to analyze the shape of sunspot cycles. This may be because of the problem of the varying cycle length. Here we circumvent this problem by using the average length of the sunspot cycle of 11.1 years (133 months) as the common cycle length for all cycles. To this end, we first resampled the monthly sunspot values so that all cycles have  the same length of 133 time steps (months). This effectively elongates or abridges the cycles to the same length. Before applying the PCA method to the resampled sunspot cycles, we standardized each individual cycle to have zero mean and unit standard deviation. This guarantees that all cycles  have the same weight in the study of their common shape.  After applying the PCA method to these resampled and standardized cycles, we reverted the cycle lengths and amplitudes to their original values. 

Standardized sunspot cycles are collected into the columns of the data matrix X, which can be decomposed as
\begin{equation}
        X = U\:D\;V^{T}  \     ,
\end{equation}
where U and V are orthogonal matrices and D a diagonal matrix 
        $D= diag\left(\lambda_{1},\lambda_{2},...,\lambda_{n}\right)$
with $\lambda_{i}$ denoting the $i^{th}$ singular value of matrix X in order of decreasing importance. The principal components are the column vectors of
\begin{equation}
  P = U\!D.
\end{equation}
The column vectors of matrix V are called empirical orthogonal functions (EOF) and they represent the weights of each principal component in the decomposition of each (standardized) cycle $X_{i}$, which can be approximated as 
\begin{equation}
        X_{i} = \sum^{N}_{j=1} \:P_{ij}\:V_{ij} \  ,
\end{equation}
where N is the number of principal components (here N=2). The variance explained by each PC is proportional to square of the corresponding singular value $\lambda_{i}$. Hence the $i^{th}$
PC explains a percentage
\begin{equation}
\frac{\lambda^{2}_{i}}{\sum^{n}_{k=1}\!\lambda^{2}_{k}} \cdot\:100\%
\end{equation}
of the variance in the data.

\section{PCA of sunspot indices for cycles 1-23}
 
\begin{figure}
\centering
\includegraphics[width=0.5\textwidth]{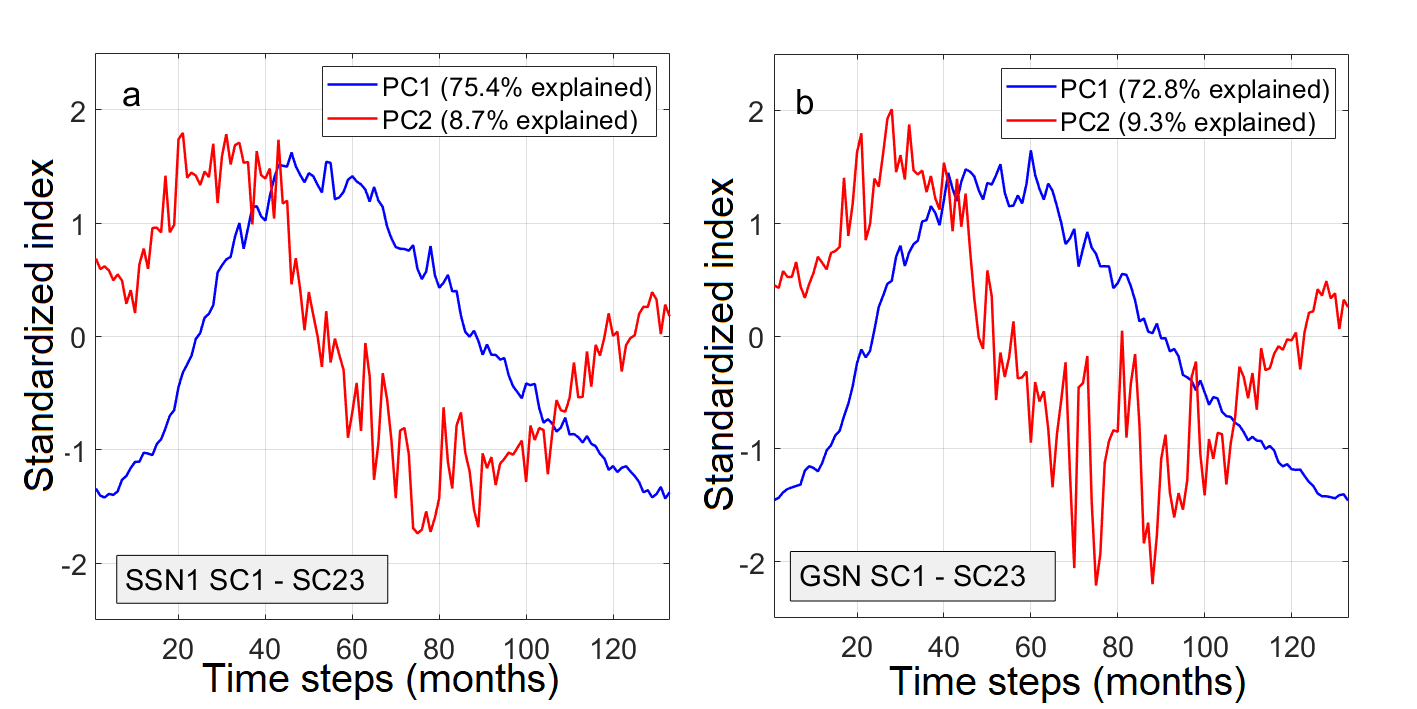}
\caption{First (blue) and second (red) principal components of a) SSN1 cycle, and b) GSN cycle.}
\label{fig:R_PCs}
\end{figure}

We first studied the common PCA shape of all the cycles SC1-SC23 using SSN1 and GSN series. Figures \ref{fig:R_PCs}a and \ref{fig:R_PCs}b show the first two main principal components of SSN1 and GSN, respectively.

As expected the first PCs (blue curve) account for most of the total variance of the sunspot cycle shape (75.4\% for SSN1 and 72.8\% for GSN). The PC1s have an ascending phase of about four to five years and a descending phase of about six to seven years long when using the common 11.1-year cycle length. The average sunspot cycle shape is, as expected, quite asymmetric. The correlation coefficient between the two PC1s is 0.993 (p < $10^{-100}$). There is some fluctuation at the top of PC1 with a small gap between local maxima, more clearly in GSN, which may be related to the so-called Gnevyshev gap (GG) \citep{Gnevyshev_1967, Gnevyshev_1977, Storini_2003, Ahluwalia_2004, Bazilevskaya_2006, Norton_2010, Du_2015}. 
 
The second PC (red curve) explains 8.7\% of the total variance for SSN1 and 9.3\% for GSN. Accordingly 84.1\% of the total variance has been captured by the first two principal components for SSN1 and 82.1\% of the total variance for GSN. The correlation coefficient of the two PC2s is 0.947 (p < $10^{-65}$). The most conspicuous feature of PC2 is its roughly sinusoidal shape, with zero crossing close to the above-mentioned Gnevyshev gap. PC2 acts to correct the shape of the cycle when the corresponding sunspot cycle differs from the average cycle shape. The main effect of the PC2 is to reduce (positive scaling for PC2) or enhance (negative scaling for PC2) the activity level of the declining phase with respect to the ascending phase of the cycle. At the same time it tends to reduce or enhance the peak after the Gnevyshev gap with respect to the peak before. 

\begin{figure}
\centering
\includegraphics[width=0.5\textwidth]{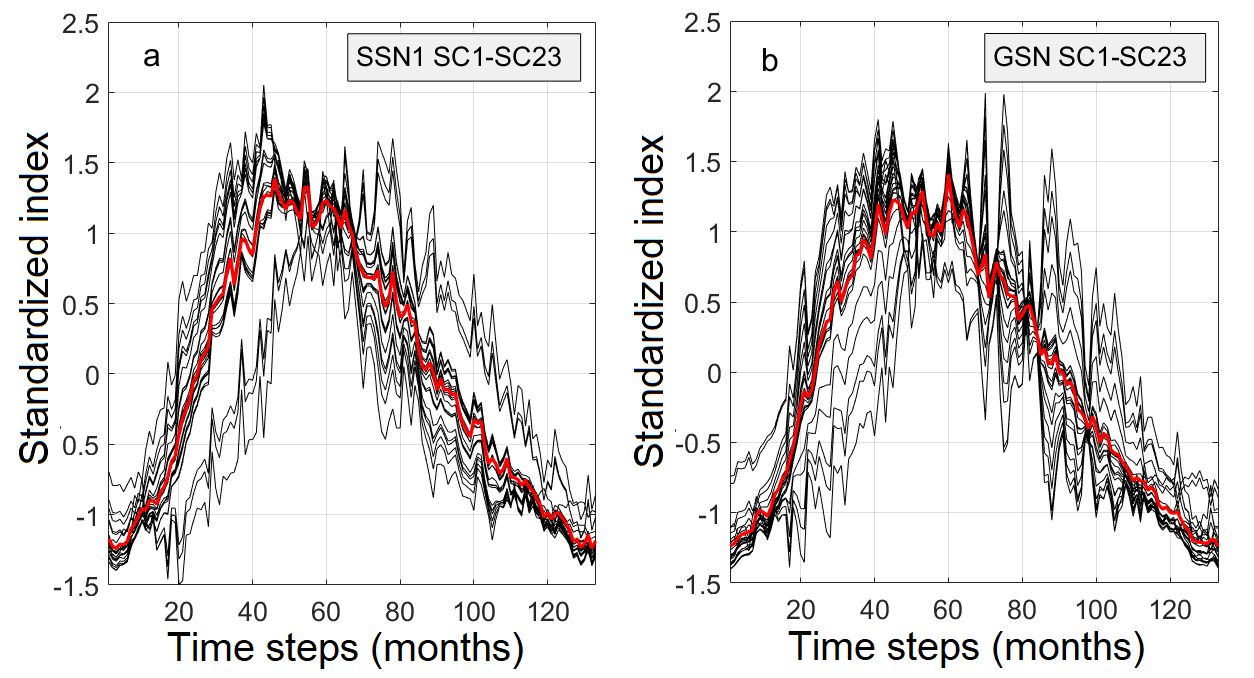}
\caption{Sums of PC1+PC2 for a) all SSN1 cycles, and b) all GSN cycles. Red curve is the corresponding average, the model cycle.}
\label{fig:R_GSN_PC1+PC2_curves}
\end{figure}

Figures \ref{fig:R_GSN_PC1+PC2_curves}a and \ref{fig:R_GSN_PC1+PC2_curves}b show the scaled sums of PC1+PC2 of all SSN1 and GSN cycles, respectively. The red curve is the corresponding average over all cycles, which is PC1. We note that, even though there is quite a large variation elsewhere, all cycles are very close to each other in the region of the Gnevyshev gap. This suggests that the Gnevyshev gap is a common fundamental property of sunspot cycles, which divides the sunspot cycle into two rather disparate parts: the ascending phase and the declining phase.
Figure \ref{fig:R_GSN_PC1+PC2_curves}a seems to suggest that the cycles would divide into two classes, the majority of more asymmetric cycles and a smaller group of more symmetric cycles. However, this separation may also result from a time shift in choosing the respective minima. Most of the symmetric cycles are from the earlier centuries when the sunspot series is less certain, which will also contribute to the differences in cycle shapes between the centuries (see later).

\begin{figure*}
\centering
\includegraphics[width=0.9\textwidth]{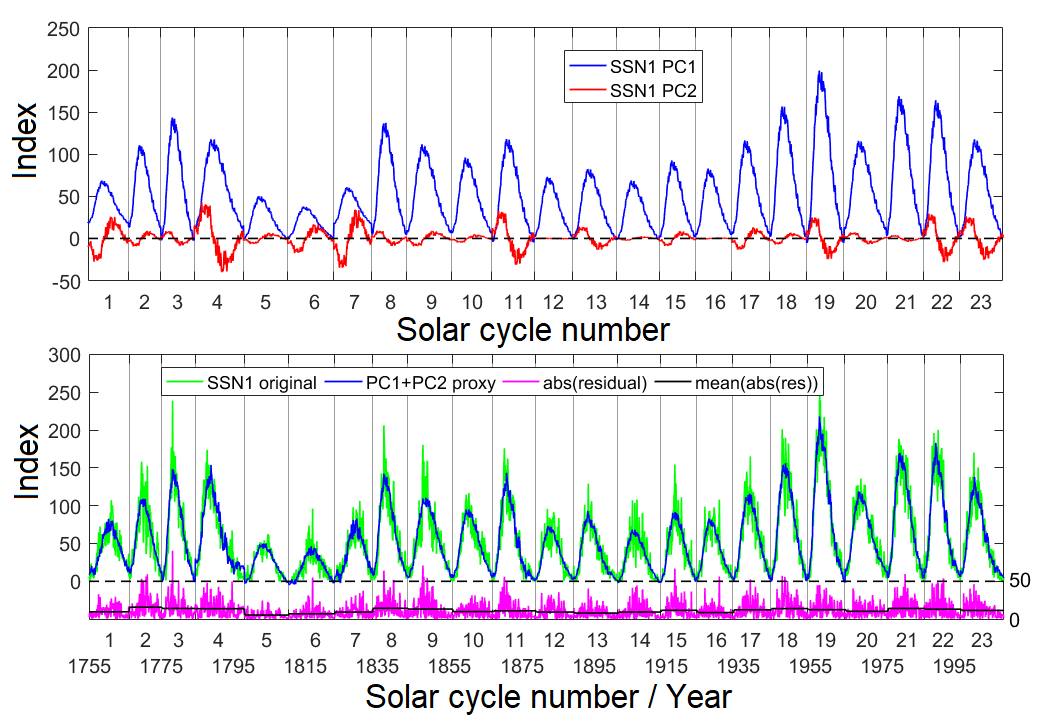}
\caption{Upper panel: PC1 (blue) and PC2 (red) time series of SSN1 for SC1-SC23. Lower panels: Original monthly SSN1 time series (green), its PC1+PC2 proxy time series (blue), and absolute value of their difference (residual) time series (magenta) with cycle means (brown). Vertical lines denote the times of sunspot minima.}
\label{fig:Rts_SSN1_PC1_PC2}
\end{figure*}

We returned each cycle back to its original length, and then back to its original amplitude by multiplying both PCs with the standard deviation of the original cycle and and adding the mean value of the original cycle to PC1. Then we concatenated the cycles to their original order and obtained the full PC1 and PC2 series, as well as the PC1+PC2 proxy sunspot series for cycles SC1-SC23. These three time series are shown in Fig. \ref{fig:Rts_SSN1_PC1_PC2} for SSN1, together with the original SSN1 series and the absolute value of the difference (residual) between SSN1 and its PC1+PC2 proxy series. We also show the average of the absolute value of the residual for each cycle with a horizontal line. The vertical lines denote the times of sunspot minima.
 
Figure \ref{fig:Rts_SSN1_PC1_PC2} shows that there are a few cycles that are defined almost solely by PC1, especially SC12, SC14, and SC16. It is interesting to note that all these "model" or representative cycles with a most typical shape are successive even cycles from around the turn of the nineteenth and twentieth centuries when the overall solar activity remained at a similar, lower than average level for a long time. Out of the recent few cycles, SC21 is  closest to the model cycle. On the other hand, there are several cycles where PC2 makes an essential contribution to the cycle shape. Cycles with a large PC2 are somewhat more common during the early centuries, probably reflecting the fact that the quality of sunspot numbers is lower in the early times. We find that SC4 and SC7 have the largest PC2 contributions, with PC2 during SC4 being largest in absolute size, while the relative fraction of PC2 is largest during the low SC7. In the early centuries, PC2 is quite large also during SC1, SC6, and SC11. Interestingly, large PC2 are also seen more recently, during the twentieth century, during SC19, SC22, and SC23. 

The lower panels of Fig. \ref{fig:Rts_SSN1_PC1_PC2} show the original SSN1, the PC1+PC2 proxy of SSN1, and the absolute value of their residual. The correlation coefficient between the original SSN1 and its PC1+PC2 proxy is 0.934 with (p < $10^{-100}$). The PC1+PC2 proxy looks very much like the smoothed curve of the original monthly SSN1 time series, although it is calculated from a sequence of 23 separate sunspot cycles. There are, however, a small number of steps in the PC1+PC2 proxy series, the largest ones around SC4 and SC7. These discontinuities are related to a large value of PC2 for these cycles.

\begin{figure}
\centering 
\includegraphics[width=0.5\textwidth]{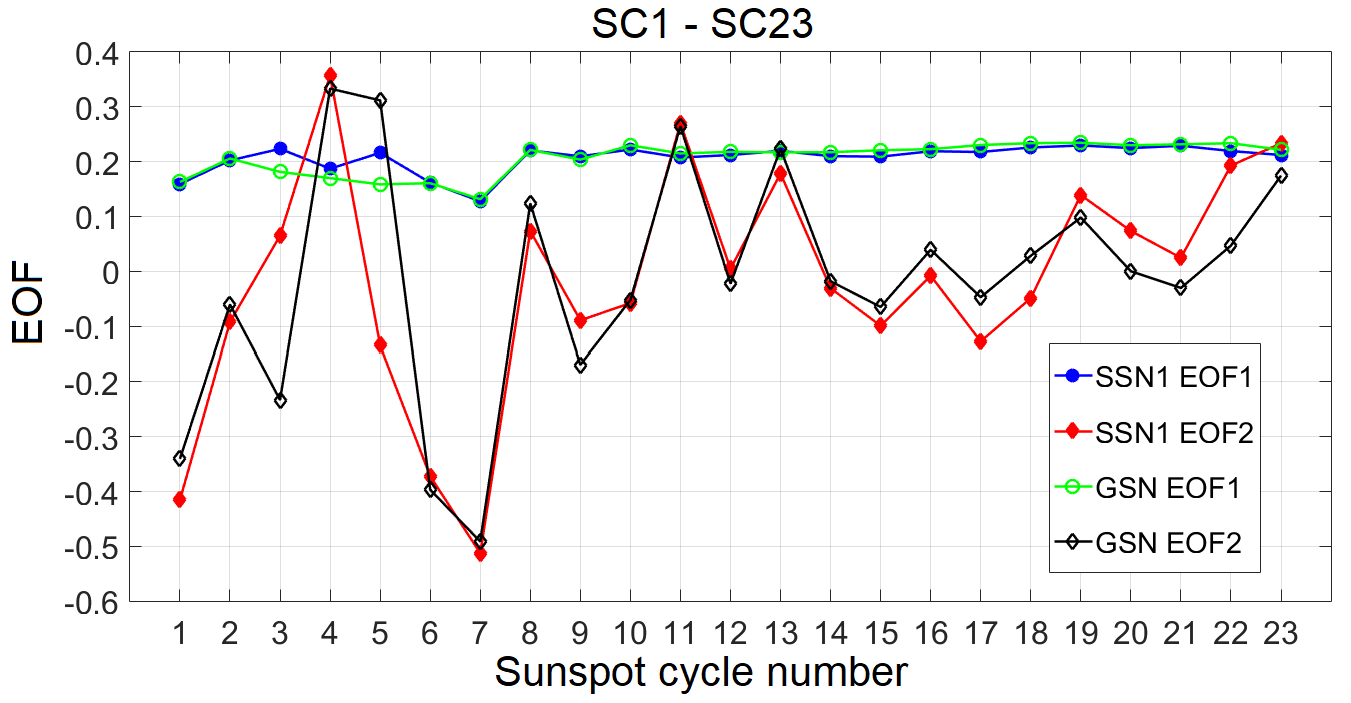}
\caption{First two EOFs of all of sunspot cycles of SSN1 and GSN.}
\label{fig:EOF_SSN1_GSN}
\end{figure}

Figure \ref{fig:EOF_SSN1_GSN} shows the two EOFs of all 23 cycles for SSN1 and GSN. The EOF1s are almost equal (about 0.2) for all cycles of both indices, except for SC1, SC6, and SC7, for both SSN1 and GSN, and for SC5 and SC6 for GSN. This means that most cycles have a roughly equal relative weight of PC1. 

The EOF2s vary considerably between the individual cycles and slightly even between SSN1 and GSN. The largest fluctuations of EOF2 values are seen in the late eighteenth and early nineteenth century. However, even then the SSN1 and GSN depict quite a similar long-term variation of EOF2, with almost all cycles (except for SC3 and SC5) attaining closely similar EOF2 values. There are altogether 12 negative and 11 positive EOF2 cycles for both SSN1 and GSN. 

During the first half of the twentieth century, when solar activity increased fairly systemically (see \ref{fig:Rts_SSN1_PC1_PC2}), EOF2s (PC2s) are mostly negative in SSN1. EOF2s are negative also for cycles 6 and 7 when cycle amplitude strongly increased after Dalton minimum. On the other hand, EOF2s are mostly positive when the following cycle is clearly smaller, like for cycles 4, 11, 13, 19, 22, and 23. A higher next cycle tends to shorten the length of the declining phase of the cycle, leading to a less asymmetric cycle than the model cycle described by PC1. The negative EOF2 (PC2) re-establishes this deviation of the cycle from the model cycle. Similarly, if the next cycle is lower, the declining phase tends to be abnormally long.

\begin{figure}
\centering
\includegraphics[width=0.5\textwidth]{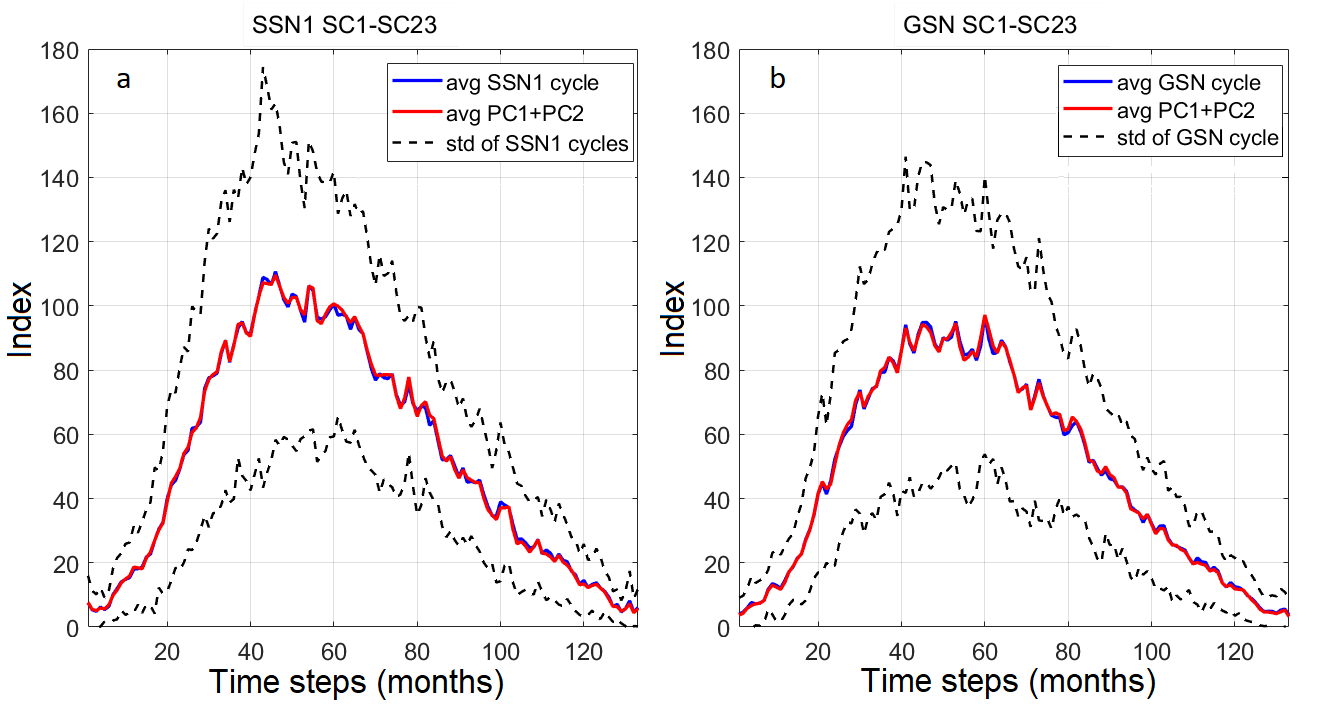}
\caption{Average cycle of a) SSN1 and b) GSN series (blue) and its PC1+PC2 proxy (red).  Standard deviations of the original cycles are shown as dashed curves.}
\label{fig:SSN1_GSN_avgs}
\end{figure}

Figure \ref{fig:SSN1_GSN_avgs} shows the averages of the original SSN1 and GSN cycles and their PC1+PC2 proxies. The standard deviations of the original SSN1 and GSN are shown as dashed curves. Although standard deviation of the cycles is quite large and about 16\% (18\%) of the total variance of SSN1 (GSN) remains unexplained by the two first PCs, the original and proxy curves are very similar in both cases. 

\begin{figure*}
\centering
\includegraphics[width=0.9\textwidth]{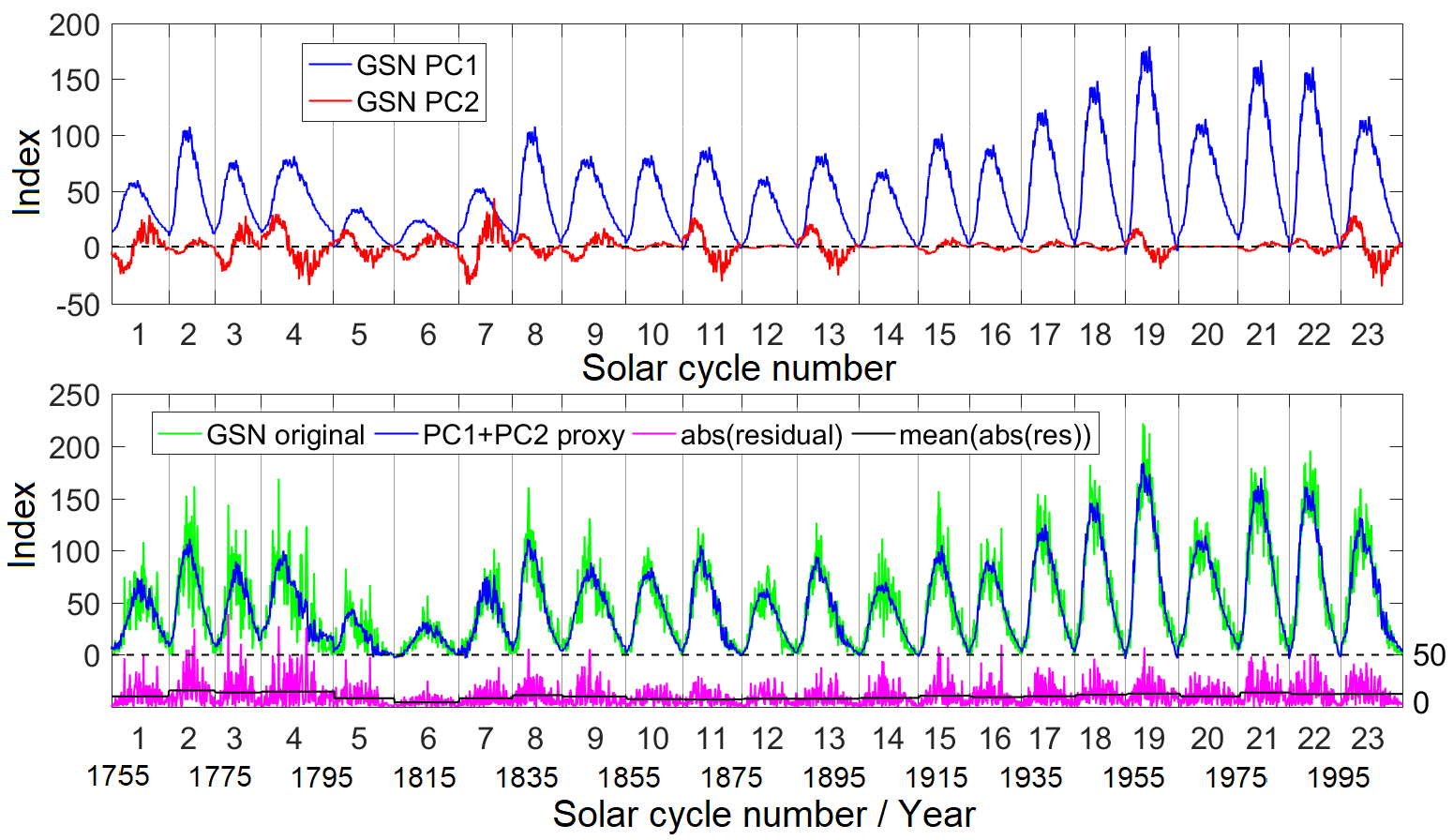}
\caption{Upper panel: PC1 (blue) and PC2 (red) time series of GSN for SC1-SC23. Lower panels: Original monthly GSN time series (green), its PC1+PC2 proxy time series (blue), and absolute value of their difference (residual) time series (magenta) with cycle means (brown). Vertical lines denote the times of GSN minima.}
\label{fig:GSNts_PC1_PC2}
\end{figure*}

Figure \ref{fig:GSNts_PC1_PC2} shows the original GSN time series and its PC1+PC2 proxy, as well as the absolute value of their residual time series. The correlation coefficient between the original GSN and its PC1+PC2 proxy is 0.936 (p < $10^{-100}$). The results are quite similar to those for SSN1, but there are some differences. It is well-known that GSN depicts a relatively lower level of sunspot activity in the eighteenth century than SSN1 \citep{Hoyt_1998, Usoskin_2003}. Thus the early cycles of the GSN series are relatively smaller than in the SSN1 series. However, more interestingly, as for SSN1, the even low cycles of roughly one hundred years ago are among the best model cycles for GSN as well. In addition, SC20 is another very good model cycle, unlike for SSN1. Figure \ref{fig:GSNts_PC1_PC2} repeats the similarity of PC2, for example, the cycle shape for GSN and SSN1, as already shown in Figure \ref{fig:EOF_SSN1_GSN}. For most cycles, the sign and even the size of PC2 is the same for GNS and SSN1.

\section{Separate PCA for the eighteenth, nineteenth and twentieth centuries}

\begin{figure}
\centering
\includegraphics[width=0.5\textwidth]{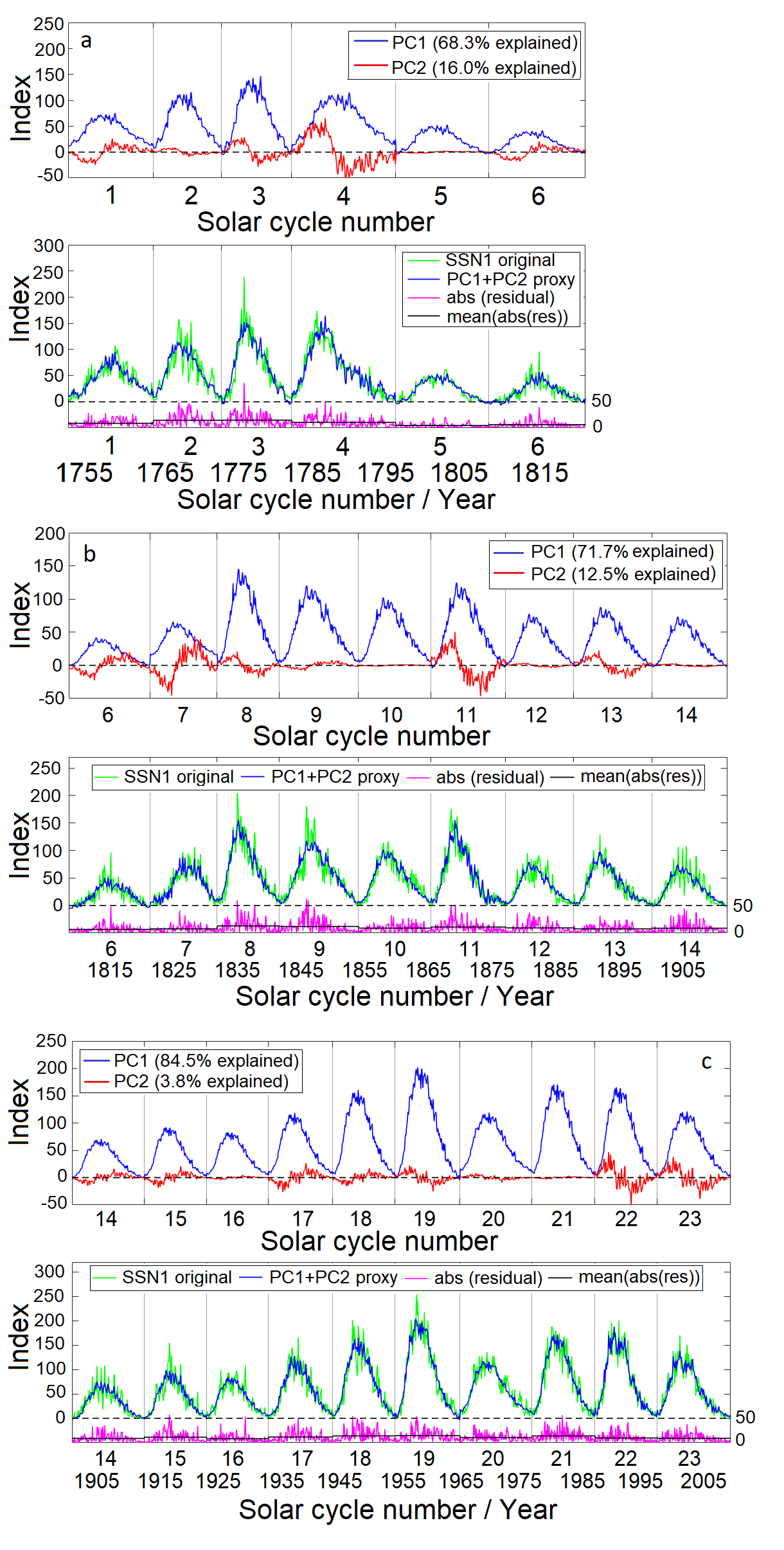}
\caption{Same information for SSN1 as in Fig. \ref{fig:Rts_SSN1_PC1_PC2}, now separately for a) eighteenth, b) nineteenth, and c) twentieth century subperiods.}
\label{fig:Sub-periods_SSN1}
\end{figure}

In order to study the properties of the cycles in different centuries, we have made separate PC analyses of SSN1 for the three centuries, including cycles SC1-SC6, SC6-SC14, and SC14-SC23 for the eighteenth, nineteenth and twentieth centuries, respectively. (The periods are slightly overlapping such that SC6 is included both in the eighteenth and nineteenth century and SC14 in the nineteenth and twentieth century subperiods.) We show the PCA results for these three subperiods in Fig. \ref{fig:Sub-periods_SSN1}. 

During each subperiod there is at least one model cycle whose PC2 is almost constant zero. During the first subperiod, SC5 is the best model cycle, with SC2 being close to it. During the nineteenth century the successive even cycles SC10, SC12, and SC14 are all very good model cycles. This suggests that the model cycle of the nineteenth century deviates only slightly from the overall model cycle of all 23 cycles depicted in Fig. \ref{fig:R_GSN_PC1+PC2_curves}. PC2 of SC6 is slightly smaller during the first than the second subperiod, indicating that SC6 resembles more closely to the model cycle of the first than the second subperiod. A similar situation is true for SC14, which is a model cycle in the second subperiod but whose PC2 is (slightly) nonzero in the third subperiod. SC16 and SC21 are the best model cycles in the twentieth century, with SC20 deviating only slightly from them.

\begin{figure}
\centering
\includegraphics[width=0.5\textwidth]{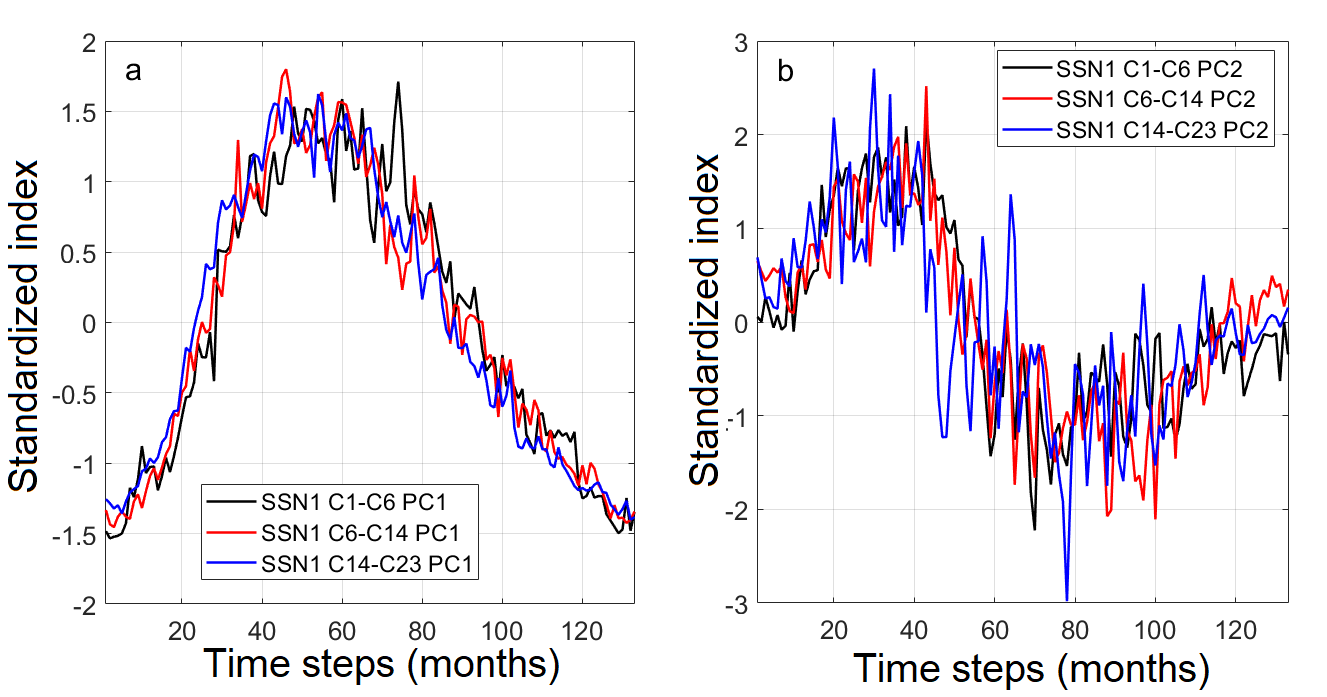}
\caption{PC1 (panel a) and PC2 (panel b) for the eighteenth (black), nineteenth (red), and twentieth (blue) century subperiods for SSN1.}
\label{fig:PCs_sub_PC1_PC2}
\end{figure}

Figure \ref{fig:PCs_sub_PC1_PC2} shows the two principal components (PC1 and PC2) for the three subperiods. PC1 explains 68.3\%, 71.7\%, and 84.5\% of the total variance of cycles in the first, second and third subperiods, respectively. The increasing percentage is at least partly due to the improved accuracy of sunspot observations in recent times, which reduces random scatter in cycle shape. We note also that PC1s of the last two centuries are closely similar, as expected based on the above discussion. These two model cycles have a slightly sharper increase and a notably earlier maximum than the PC1 of the eighteenth century, which has a  dubiously sharp maximum around the 70-75 month. The latter reflects the problem of the cycles in the eighteenth century mentioned above and depicted in Fig. \ref{fig:R_GSN_PC1+PC2_curves}. Accordingly the correlation coefficient between the first and second subperiod PC1s (0.976, p < $10^{-87}$) is better than between the first and third (0.951, p < $10^{-67}$), or second and third subperiod PC1s (0.957, p < $10^{-71}$).

PC2 accounts for 16.0\%, 12.5\%, and 3.8\% of the variance of cycles in the three subperiods, respectively. PC1 and PC2 together explain 84.3\%, 84.2\%, and 88.3\% of variance in the three subperiods; this is almost the same amount for all centuries. All PC2s show the same sinusoidal shape (see Fig. \ref{fig:PCs_sub_PC1_PC2}b) but their mutual similarity is fairly small. The correlation coefficients of the first and second, first and third, and second and third subperiod PC2s are 0.849 (p < $10^{-37}$), 0.686 (p<$10^{-19}$), and 0.682 (p<$10^{-18}$), respectively.

\begin{figure}
\centering
\includegraphics[width=0.5\textwidth]{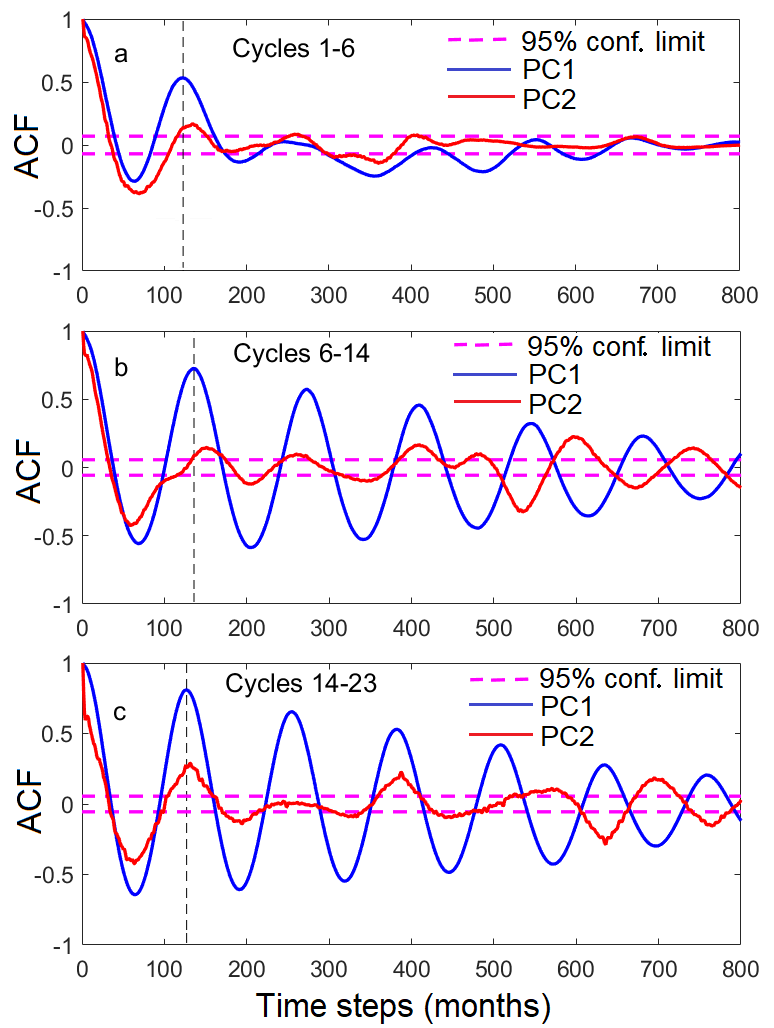}
\caption{Autocorrelation functions of the PC1 time series (blue) and the PC2 time series (red) of a) eighteenth, b) nineteenth, and c) twentieth century subperiods, respectively.}
\label{fig:ACFs_sub_PC1_PC2}
\end{figure}

Figure \ref{fig:ACFs_sub_PC1_PC2} shows the autocorrelation functions (ACF) of the PC1 and PC2 proxy series for the three subperiods. The horizontal dashed lines show 95\% confidence limits of the ACFs. (We use here the 95\% CL as a limit for statistical significance). The first maximum of the PC1 ACFs gives the mean length of model sunspot cycle of the respective subperiod. Figure \ref{fig:ACFs_sub_PC1_PC2} shows that the mean lengths are 123, 137, and 127 months for the the three subperiods, respectively. Corresponding cycle lengths, when calculated from the original SSN1 subperiod time series, are 123, 136, and 127 months. We also note  that the PC1 ACF of the first subperiod reduces to an insignificant level soon after the first maximum. This is probably due to the mutually rather different lengths of cycles of that period, in particular to the extremely long SC4.

PC2 is also somewhat periodic, although its significance is, especially beyond the first maximum, somewhat marginal. While the first maximum of the PC2 ACF of the third subperiod is found roughly at the same time as the corresponding PC1 maximum, the PC2 maximum of the first and second subperiods (at 129 and 152 months, respectively) have a slightly longer lag than the corresponding PC1 maximum. After three to four cycles the PC2 experiences a phase shift, whereafter PC1 and PC2 are in antiphase. This is significant for the PC2 of the second subperiod after three cycles and for the third subperiod after four cycles. 
These results show that while there is no systematic long-term evolution of cycle shape asymmetry in the eighteenth century, the nineteenth and especially the twentieth centuries have a sequence of a few (3-4) cycles with increasing height and positive asymmetry (PC2) and, thereafter, cycles of decreasing height and negative asymmetry.

\begin{figure}
\centering
\includegraphics[width=0.5\textwidth]{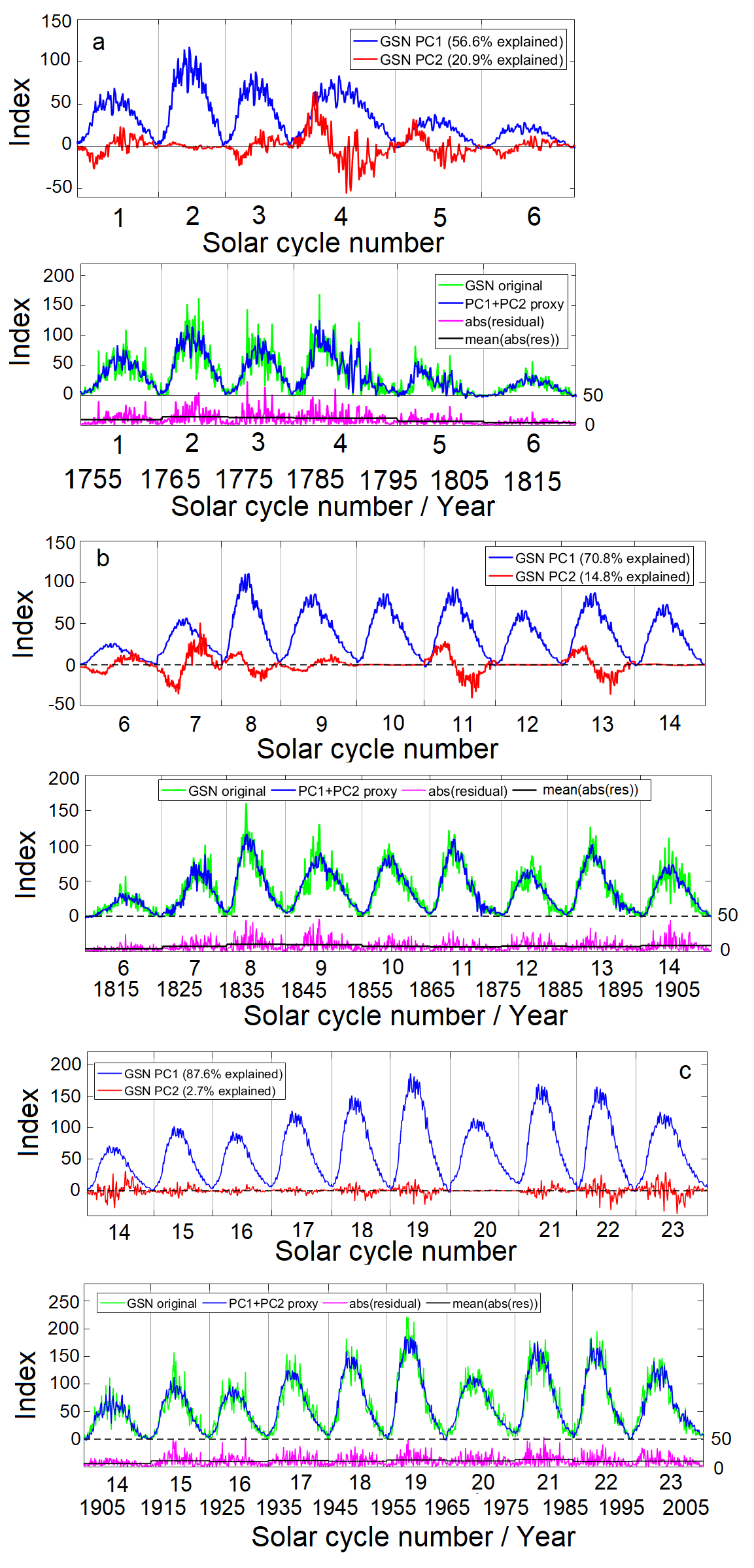}
\caption{Same information as in Fig. \ref{fig:Sub-periods_SSN1} but for GSN.}
\label{fig:Sub-periods_GSN}
\end{figure}

Figure \ref{fig:Sub-periods_GSN} shows the PC1 and PC2 time series for the three subperiods of GSN. As for SSN1,  cycle SC4 needs a very strong PC2 correction term, but now only SC2 is a model cycle of the first subperiod. Cycles 10, 12, and 14 are the model cycles of the second subperiod for GSN, as for SSN1. However, SC20 is the only clear model cycle in the twentieth century for GSN, rather than SC21 or SC16. This suggests that the shape of the model cycle for GSN is roughly the same as for SSN1 in the nineteenth century subperiod but somewhat different than SSN1 in eighteenth and twentieth centuries. 

\begin{figure}
\centering
\includegraphics[width=0.5\textwidth]{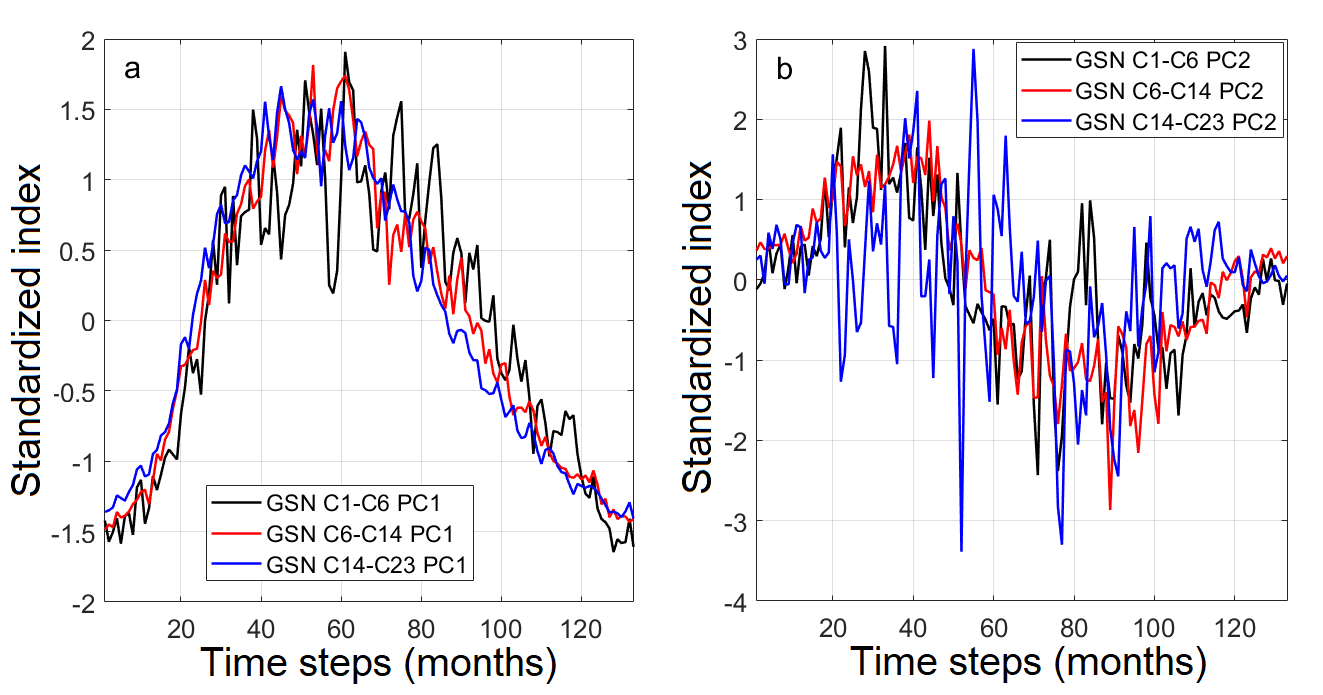}
\caption{PC1 (panel a) and PC2 (panel b) for the eighteenth (black), nineteenth (red), and twentieth (blue) century subperiods for GSN.}
\label{fig:GSN_PCs_sub_PC1_PC2}
\end{figure}

Figure \ref{fig:GSN_PCs_sub_PC1_PC2} shows the two principal components of GSN for each subperiod. PC1 explains 56.6\%, 70.8\%, and 87.6\% and PC2 20.9\%, 14.8\%, and 2.7\% of the total variance of the first, second, and third subperiods, respectively. The percentage of PC1 during the eighteenth century is much smaller for GSN than for SSN1. This may be due to the larger number of data gaps in the GSN (than SSN1) series at this time. This also reduces the total amount of variability explained by PC1 and PC2 together for the first subperiod to be lower, 77.5\%, than for the second (85.6\%) and the third (90.3\%) subperiod. The total variability explained by PC1 and PC2 for SSN1 and GSN are roughly the same for the second and third subperiod. The shape of PC1 for the first subperiod again differs from the other subperiods and depicts similar dubious variability in the declining phase, as for SSN1. Somewhat surprisingly, the PC2 of the twentieth century is least sinusoidal and is quite different from the PC2 of the two previous centuries. The correlation coefficients for GSN PC1 and PC2 between the first and second subperiods are 0.918 (p < $10^{-53}$) and 0.784 (p < $10^{-28}$), between the first and third subperiods 0.893 (p < $10^{-46}$) and 0.316 (p < 0.00021), and between the first and third subperiods 0.976 (p < $10^{-87}$) and 0.424 (p < $10^{-6}$), respectively.

\section{Waldmeier rules}

\begin{figure}
\centering
\includegraphics[width=0.5\textwidth]{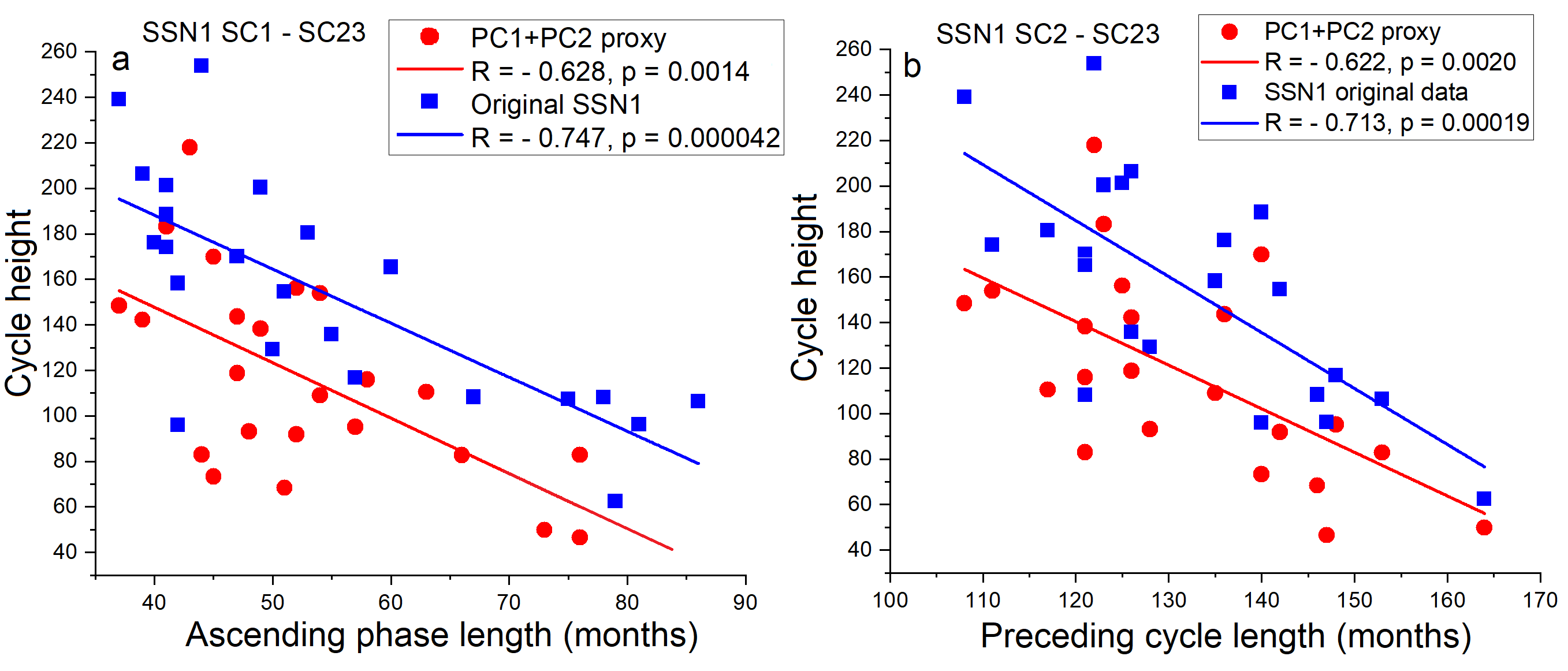}
\caption{Cycle height as a function of a) the length of ascending phase of the same cycle, b) of the length of preceding cycle for SSN1.}
\label{fig:Waldmeier_effect_SSN1}
\end{figure}

In order to test the two Waldmeier rules mentioned above, we plotted the height (the absolute maxima) of each SSN1 sunspot cycle as a function of the length of the ascending phase of the same cycle in months in Fig. \ref{fig:Waldmeier_effect_SSN1}a, and as a function of the length of the preceding sunspot cycle in months in Fig. \ref{fig:Waldmeier_effect_SSN1}b, using both the original sunspot cycles and their PC1+PC2 proxy cycles. 

The regressions in Fig. \ref{fig:Waldmeier_effect_SSN1}a have correlation coefficients of -0.747 (p = 0.000042) and -0.628 (p = 0.0014) for original cycles and PC1+PC2 proxies, respectively, thus verifying the significant anticorrelation between cycle height and the length of the ascending phase (Waldmeier rule 1). In the proxy, the cycle maxima are more smoothed, which lowers and spreads the cycle maxima and may be the reason for the somewhat weaker anticorrelation. Similarly, Fig. \ref{fig:Waldmeier_effect_SSN1}b verifies the significant anticorrelation between cycle height and the length of the preceding cycle for both original and proxy cycles, with correlation coefficients of -0.713 (p=0.00019) and -0.622 (p=0.0020), respectively, thus confirming Waldmeier rule 2.

\begin{figure}
\centering
\includegraphics[width=0.5\textwidth]{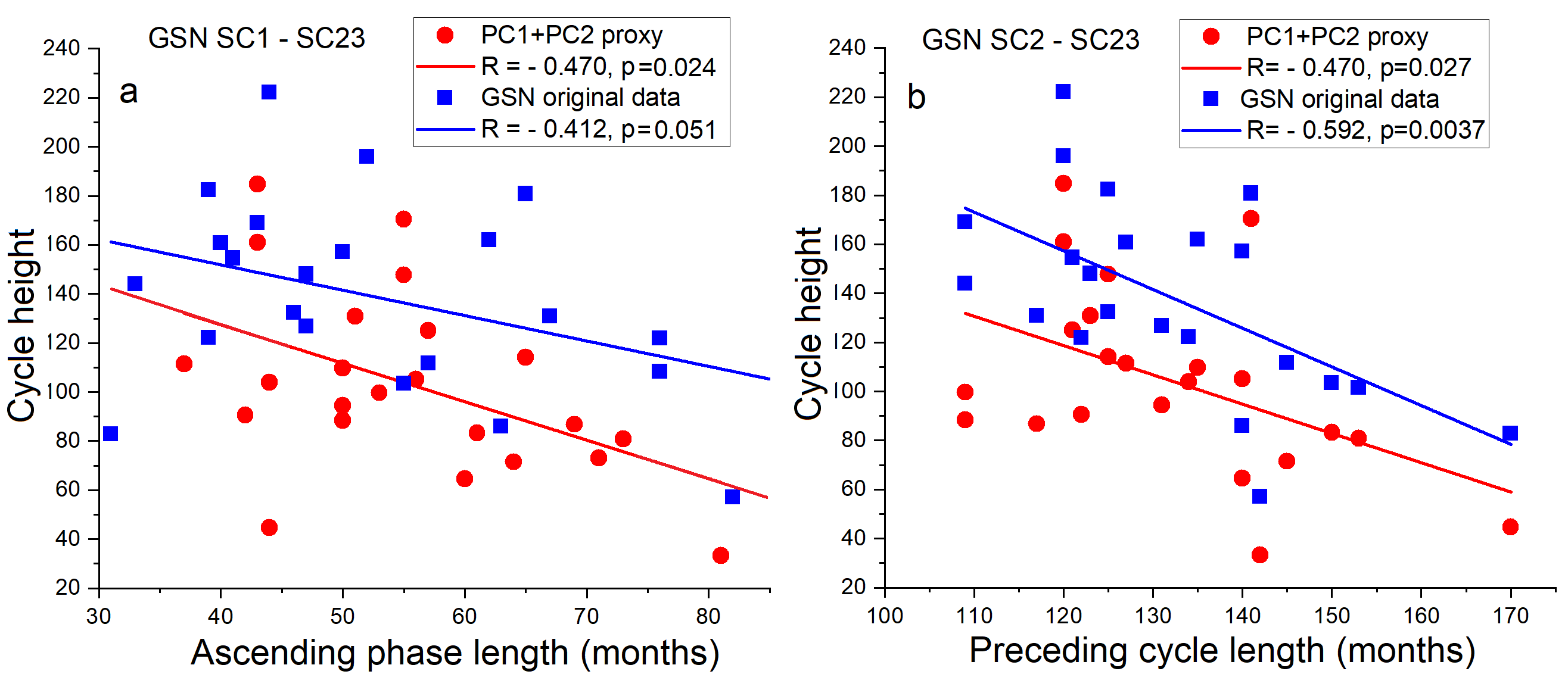}
\caption{Cycle height as a function of a) the length of ascending phase of the same cycle, b) of the length of preceding cycle for GSN.}
\label{fig:Waldmeier_effect_GSN}
\end{figure}

The same analysis for GSN cycles is shown in Fig. \ref{fig:Waldmeier_effect_GSN}. Waldmeier rule 2 (Fig. \ref{fig:Waldmeier_effect_GSN}b) is verified for both original and proxy cycles, although their significance is somewhat weaker than for SSN1. This result is in line with the findings of \cite{Aparicio_2012}  that anticorrelation is  worse for GSN than for SSN.
For GSN, regression is weaker for the proxy series than the original, as for SSN1. However, Waldmeier rule 1 (Fig. \ref{fig:Waldmeier_effect_GSN}a) is not significant for the original GSN record, but becomes marginally significant for the proxy series. This is probably due to the fact that the proxy series tends to relocate the cycle maxima to different times for cycles whose maxima are "abnormally" located. This is the case particularly for the cycles of the eighteenth century, where the difference between the two maxima is large for several cycles. In most cases, for example, for SC1, SC2, and SC6 (see Fig. \ref{fig:Sub-periods_GSN}a, lower panel), the proxy cycle maximum is at an earlier time than the original series, but the opposite situation can also be found (SC3).
This feature becomes more significant for GSN than the effect of smoothing which, as discussed above, tends to reduce significant correlations for SSN1.

\begin{figure}
\centering
\includegraphics[width=0.5\textwidth]{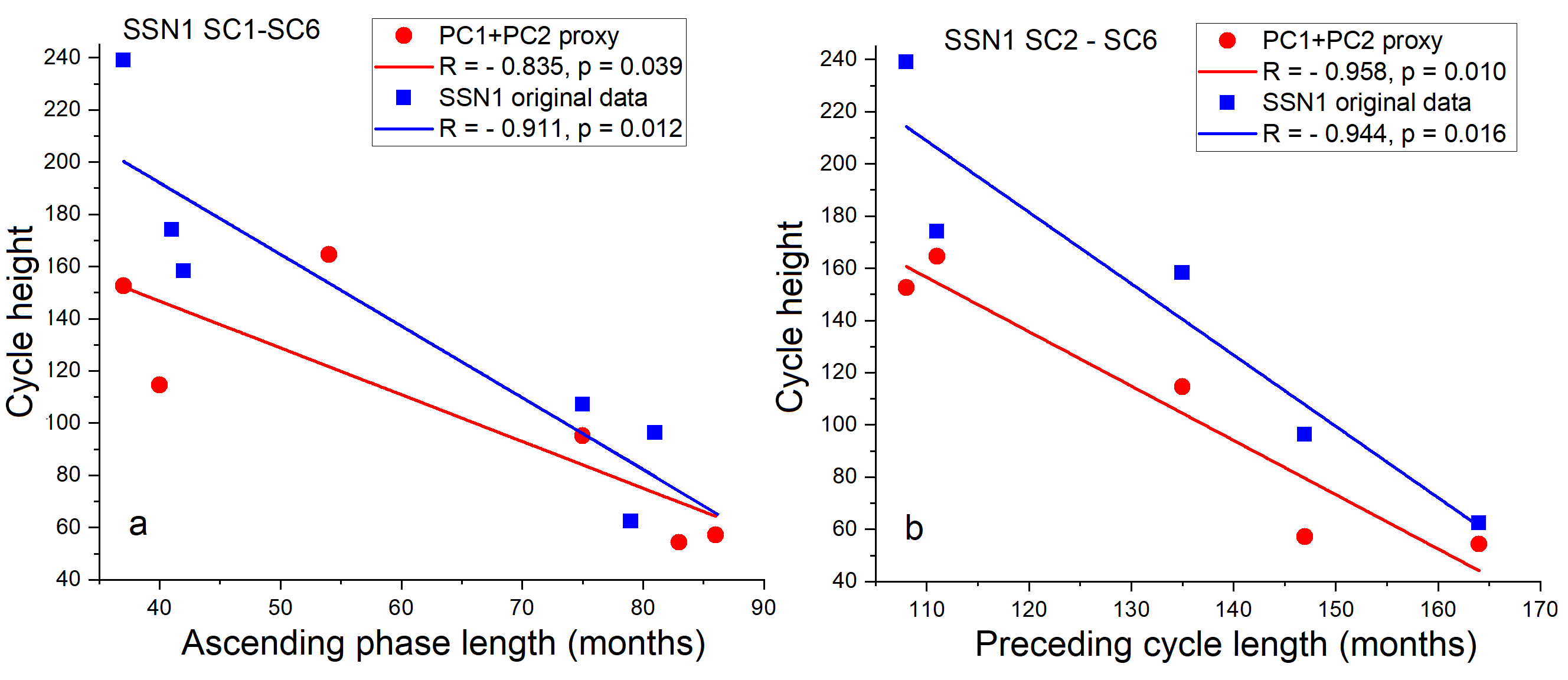}
\caption{Cycle height as a function of a) the length of ascending phase of the same cycle, b) of the length of preceding cycle for eighteenth century cycles of SSN1.}
\label{fig:Waldmeier_effect_SC1_SC6}
\end{figure}

\begin{figure}
\centering
\includegraphics[width=0.5\textwidth]{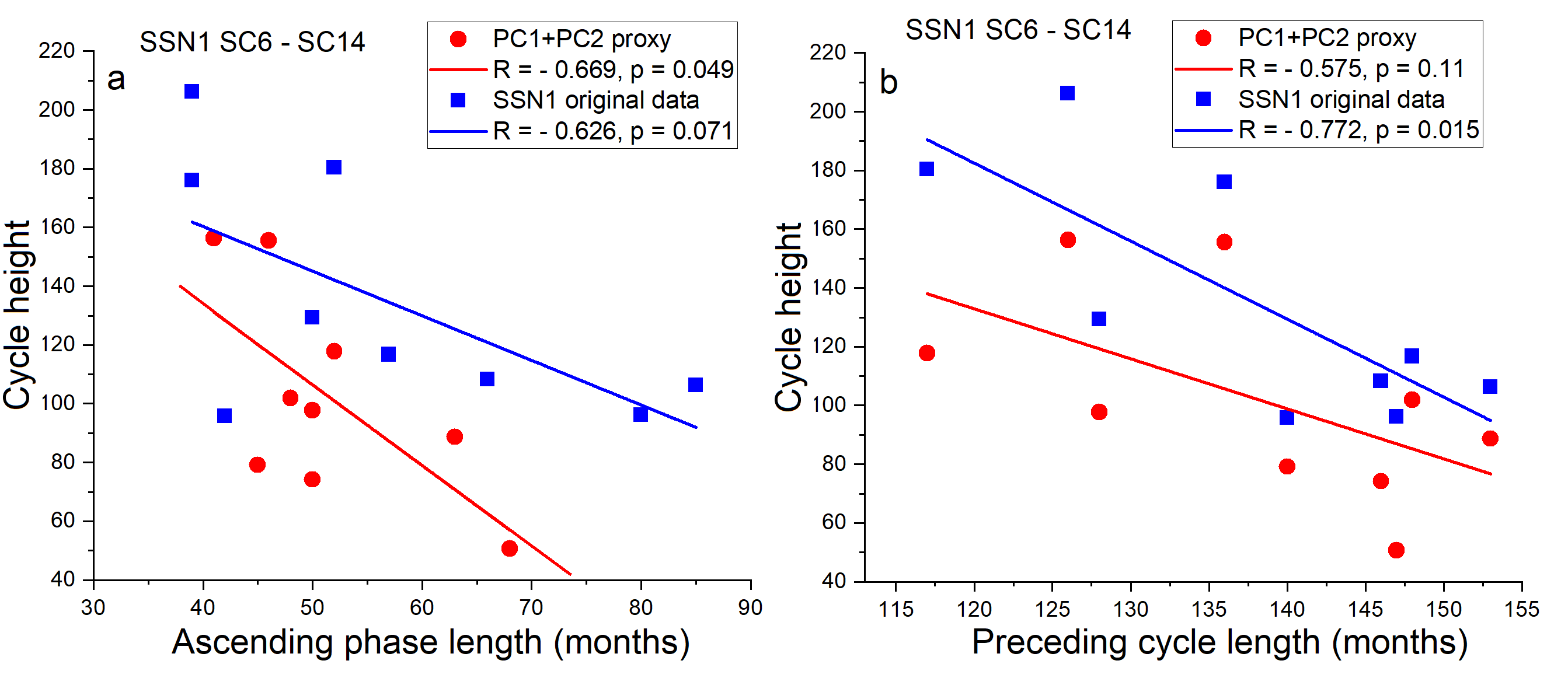}
\caption{Cycle height as a function of a) the length of ascending phase of the same cycle, b) of the length of preceding cycle for nineteenth century cycles of SSN1.}
\label{fig:Waldmeier_effect_SC6_SC14}
\end{figure}

\begin{figure}
\centering
\includegraphics[width=0.5\textwidth]{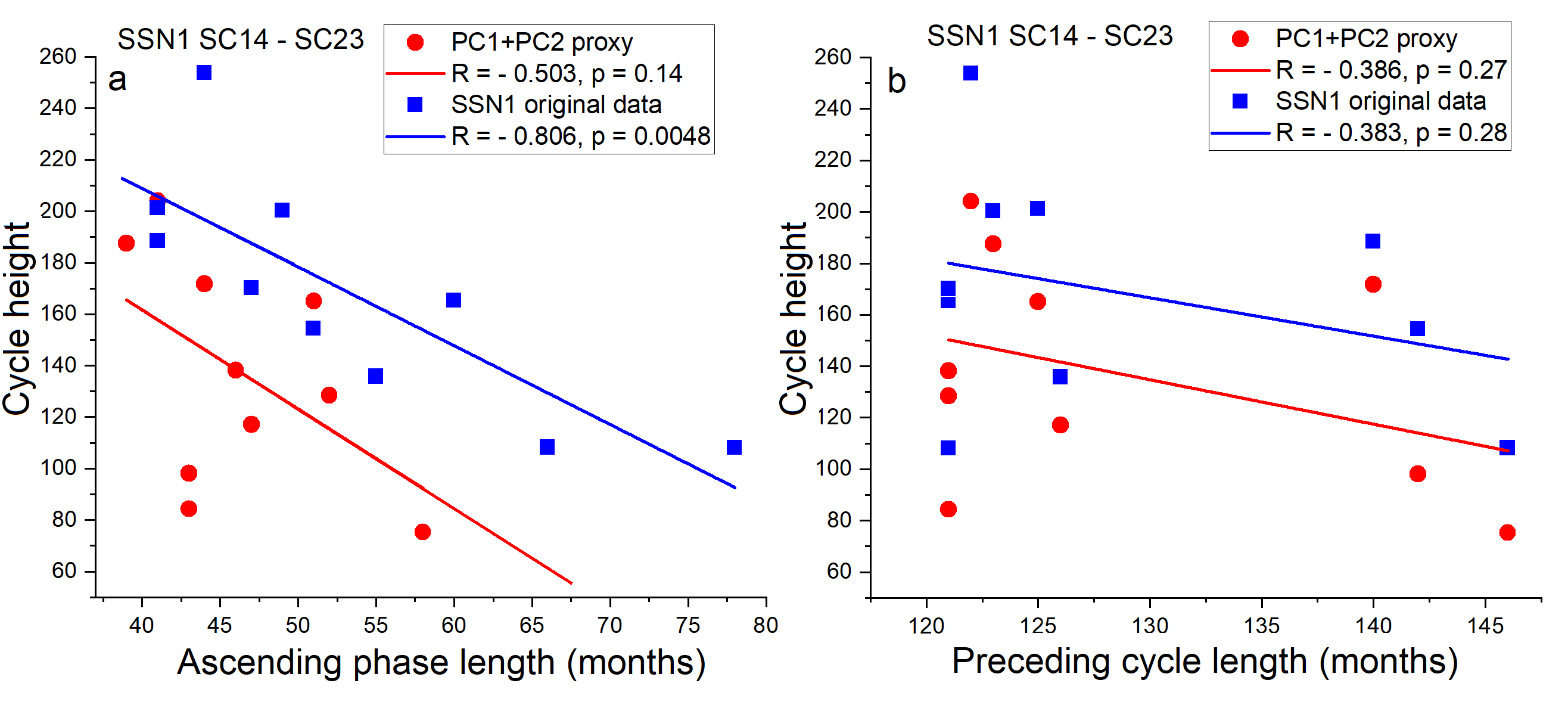}
\caption{Cycle height as a function of a) the length of ascending phase of the same cycle, b) of the length of preceding cycle for twentieth century cycles of SSN1.}
\label{fig:Waldmeier_effect_SC14_SC23}
\end{figure}

Figures \ref{fig:Waldmeier_effect_SC1_SC6}, \ref{fig:Waldmeier_effect_SC6_SC14}, and \ref{fig:Waldmeier_effect_SC14_SC23} study the validity of the Waldmeier rules separately in the eighteenth, nineteenth, and twentieth century, respectively. Figure \ref{fig:Waldmeier_effect_SC1_SC6} confirms the two Waldmeier rules for the eighteenth century for both the original and proxy series. The correlation coefficient is high in all the four cases and even though the number of points is small, the anticorrelations are statistically significant. 

Surprisingly, correlations are generally weaker and mostly insignificant during the nineteenth and twentieth centuries. Waldmeier rule 1 is valid in the original SSN1 series in the twentieth century, but not in the nineteenth century nor in the proxy series in either century. On the other hand, Waldmeier rule 2 is valid in the original SSN1 series in the nineteenth century, but not in the twentieth century nor in the proxy series in either century. There are no large differences in the timing of SSN1 cycle maxima between the original and proxy series in the eighteenth century (see Fig. \ref{fig:Sub-periods_SSN1}a, lower panel), contrary to the GSN series (see above discussion). However, there are a few cycles in the nineteenth (especially SC7) and twentieth centuries (SC16) with considerable differences (see Figs. \ref{fig:Sub-periods_SSN1}b and \ref{fig:Sub-periods_SSN1}c, lower panels). This explains why proxy series typically lead to weaker correlations in both Waldmeier rules in the nineteenth and twentieth centuries. Similar analysis for GSN in the eighteenth, nineteenth, and twentieth centuries shows mostly insignificant correlations (not shown). Waldmeier rule 1 is valid in the nineteenth century in both original and proxy series. All other five regressions are insignificant.

\section{Discussion and conclusions}

In this paper we have studied the Z\"urich sunspot number series (SSN1, original version) and the group sunspot number (GSN) series by \cite{Hoyt_1998} for the sunspot cycles 1-23 using  principal component analysis. We used the standard cycle minima for the SSN1 data \citep{NGDC_2013} and defined the minima for GSN using the 13-month Gleissberg filter. We  resampled the monthly sunspot values so that all cycles have the same mean length of 133 months. Before applying the PCA method, we  standardized each individual cycle to have zero mean and unit standard deviation. In this way the cycle amplitudes do not affect their common shape. The first two PCA components explain 84.1\% and 82.1\% of the total variance of SSN and GSN, respectively. 

PC1 gives the average shape of the solar cycle, the so-called model cycle. We found that there are some cycles that are almost solely defined by PC1, thus representing perfect model cycles. Using the whole interval of 23 SSN1 cycles, the best model cycles are SC12, SC14, and SC16 (with SC21 deviating only slightly from these). These three cycles are successive even cycles from around the turn of the nineteenth and twentieth centuries, when the overall solar activity remained at a rather low level for a long time. They are roughly equally high, and all of them lower than the odd cycles on their either side. However, their lengths deviate quite considerably from 10.1 years to 11.8 years. If their heights were exactly the same, Waldmeier rule 1 (anticorrelation between cycle length and its ascending phase) would naturally be violated. However, the longest cycle SC14 is also lowest, making the rule possible.

We  also studied the cycle shapes in three different centuries separately, making PC analyses for cycles SC1-SC6, SC6-SC14, and SC14-SC23, which represented the eighteenth, nineteenth, and twentieth centuries, respectively. The best SSN1 model cycle for the eighteenth century alone is SC5. For the nineteenth century, the successive even cycles, SC10, SC12, and SC14 form the best model cycles. For the twentieth century, SC16 and SC21 are the model cycles, slightly better than SC20.

For the whole GSN series, almost the same cycles (SC12, SC14, and SC20) as for SSN1 are found to be the best model cycles. For the three centuries of GSN separately, SC2 is the model cycle for the eighteenth century, SC10, SC12, and SC14 for the nineteenth century, and SC20 for the twentieth century. Accordingly, we find that even cycles are almost exclusively the model cycles in almost all of the eight different analyses. This is probably related to the Gnevyshev-Ohl rule, which is an expression of the general 22-year variation of cycle amplitudes and intensities \citep{Mursula_2001}, according to which even cycles are, on  average, about 10-15\% lower than odd cycles. Since a high number of activations increase the possibility of random shape distortion, lower even cycles have a higher tendency of maintaining the regular cycle evolution and, thereby, the model cycle shape.

The second PC component makes the leading correction of each individual cycle from the model cycle. Cycles 4 and 7 have the largest PC2 contributions both in SSN1 and GSN, with the relative share of PC2 being largest during SC7. These two cycles have been known for a long time to have abnormal lengths and/or shapes. The large positive PC2 (sign defined in Fig. \ref{fig:R_PCs}) of SC4 tries to reduce the excessive activity of the declining phase of the PC1 of this longest-ever cycle, while the negative PC2 of SC7 aims to recover the exceptional activity of the declining phase of this cycle with respect to the model cycle. Other cycles with a large PC2, that is, cycles that largely deviate from model cycle shape, are SC1, SC11, SC19, SC22, and SC23 in SSN1 and SC1, SC3, SC11, SC13, SC19, and SC23 in GSN. We note that almost all of these cycles are odd cycles. 

In addition to the effect of errors to the PC2 of problematic cycles, especially in the eighteenth century, the PC2 component of the individual cycles can largely be explained by the effect of the following cycle. A higher next cycle tends to shorten the length of the declining phase of the cycle, leading to a less asymmetric cycle with a higher activity in the declining phase than in the model cycle. The ensuing negative PC2 aims to re-establish the higher activity of the cycle compared to the model cycle. Similarly, if the next cycle is lower, the declining phase of the cycle tends to be abnormally long and the PC2 is positive. For example, during the first (latter) half of the twentieth century, when solar activity increases (decreases) fairly systematically, the SSN1 PC2s are mostly negative (positive, respectively; see \ref{fig:Rts_SSN1_PC1_PC2}). These results confirm and clarify the earlier suggestion of the effect of cycle overlap on cycle shape and length \citep{Hathaway_1994, Solanki_2002}.

We have used  autocorrelation to study the possible recurrence structure of the PC2 time series. Our results show that, while there is no systematic long-term evolution of cycle shape asymmetry in the eighteenth century, the PC2 series of the nineteenth century and, especially, the 20th century, has a tendency to recur after three to four cycles. After the corresponding lag, the PC2 autocorrelation function reverses its phase, the following maxima remaining still significant. Accordingly, in the two last centuries, there was a sequence of three to four cycles with increasing height and positive PC2 and, thereafter, cycles of decreasing height and negative PC2. This verifies that the cycle shapes are indeed in agreement with the centennial evolution of solar activity during these two centuries. On the other hand, no similar agreement is found in the eighteenth century, probably because of errors in sunspot number series at that time.

Our study supports the existence of the Gnevyshev gap and gives it a new interpretation. In addition to the traditional view of being a local depression of solar activity, the Gnevyshev gap is a separatrix that divides the solar cycle into two parts (the extended ascending phase and the declining phase) whose relative activity determine the cycle asymmetry. The latter interpretation is validated by the result that the Gnevyshev gap is the zero value time of the PC2, which determines the asymmetry of the cycle. The time of the the Gnevyshev gap is 45-55 months after the start of the nominal cycle length, that is, approximately 33-42\% into the cycle after its start.

We have also tested the two Waldmeier rules, that is, the anticorrelation between cycle height and the length of its ascending phase (rule 1), and the anticorrelation between cycle height and the length of the preceding cycle (rule 2), for the two original sunspot series (SSN1 and GSN) and their PCA proxy series. We find that both Waldmeier rules are valid for SSN1 for the whole sequence of 23 cycles. Waldmeier rule 2 is also valid for GSN series (both original and proxy), although its significance is weaker than for SSN1. However, Waldmeier rule 1 is significant only for the proxy series, not for the original GSN series. The proxy series improves correlation by relocating the cycle maxima to more "standard" times for cycles whose maxima are "abnormally" located. This is important only for  cycles of the eighteenth century, where the difference between these two maxima is large for several GSN cycles. When calculated separately for the last three centuries, Waldmeier rule 1 is valid for SSN1 in the eighteenth and twentieth centuries, and rule 2 in the eighteenth and nineteenth centuries. For GSN, only Waldmeier rule 1 in the nineteenth century is valid, but all other regressions remain insignificant. 

Finally, we note that the results based on the PC method, such as those presented here, offer a new tool for the analysis of the  phase space evolution and chaotic and other non-linear properties of sunspot series and solar cycles either observed or produced, for example, by solar dynamo models.

\newpage

\begin{acknowledgements}

We acknowledge financial support by the Academy of Finland to the ReSoLVE Centre of Excellence (project no. 272157). The sunspot data were obtained from WDC-SILSO, Royal Observatory of Belgium, Brussels (http://sidc.be/silso/). The dates of the cycle minima and the lengths of the SSN cycles were obtained from  the National Geophysical Data Center (NGDC), Boulder, Colorado, USA (ftp.ngdc.noaa.gov).

\end{acknowledgements}

\newpage

\bibliographystyle{aa}

\bibliography{references_JT}

\end{document}